\documentclass[a4paper,11pt]{article}
\usepackage{pos}

\def\R{{\mathbb R}}  
 \def\one{\mbox{1 \kern-.59em {\rm l}}}

\def\cM{{\cal M}}

\def\l{\lambda}

\newcommand{\eq}[1]{(\ref{#1})}

\def\nn{\nonumber} 

\newcommand{\qm}[1]{``#1''}

\title{On the propagation across the big bounce in an open quantum FLRW cosmology}

\author*[a]{Emmanuele Battista}
\author[a]{Harold C. Steinacker}

\affiliation[a]{Faculty of Physics, University of Vienna\\
 Boltzmanngasse 5, A-1090 Vienna, Austria}

\emailAdd{emmanuele.battista@univie.ac.at}
\emailAdd{harold.steinacker@univie.ac.at}

\abstract{Recently, solutions of the Ishibashi, Kawai, Kitazawa and Tsuchiya matrix theory  have been found, which can be interpreted as 3+1-dimensional quantum geometries describing an effective Friedmann-Lema\^{i}tre-Robertson-Walker cosmology with a big bounce. In this paper, we examine the propagation of a  scalar field in an open Friedmann-Lema\^{i}tre-Robertson-Walker spacetime arising within this framework. The paper is divided into two parts. In the first one, we perform a classical investigation by resorting to general-relativity tools where we show that  both massless and massive non-interacting particles  can  travel across the big bounce. In the second part,  we  evaluate the scalar field propagator by means of quantum-field-theory techniques. This analysis reveals  that in the late-time regime the scalar propagator resembles the standard Feynman propagator of flat Minkowski space, whereas for early times it gives rise to a well-defined correlation between two points on opposite sheets of the spacetime. The paper is based on Ref. \cite{Battista:2022hqn}.}

\FullConference{%
  Corfu Summer Institute 2022 "School and Workshops on Elementary Particle Physics and Gravity",\\
  28 August - 1 October, 2022\\
  Corfu, Greece}

\begin{document}
\maketitle

\section{Introduction}

Standard cosmology relies on two theoretical frameworks: the standard model of particle physics and general relativity \cite{Peebles1994}. Despite its success, the standard cosmological paradigm suffers from a series of issues such as  the cosmological horizon, the flatness problem, the baryon asymmetry,  the dark energy and dark matter puzzles, and the initial big-bang singularity. The latter problem can be overcome by resorting to  nonsingular bouncing cosmological models, where the big bang is replaced by a big bounce (BB) as the universe goes from a contracting era to an expanding epoch, see e.g. Refs.  \cite{Klinkhamer:2019dzc,Klinkhamer2019b,Wang2021a,Battista2020a,Steinacker:2017vqw,Steinacker:2017bhb,Sperling:2019xar,Klinkhamer2020a,Odintsov2019,Odintsov2020a,Odintsov2020b,Easson2011,Ijjas2016,Ijjas2016b,Cai2012,Corichi2007,Bojowald2008,Bojowald2009,Wilson-Ewing2012,Veneziano2003,Gasperini2007,Novello2008,Cai2014,Battefeld2014,Banerjee2022,Agrawal2022}.

Recently, bouncing cosmological models have been found in the context of the Ishibashi, Kawai, Kitazawa and Tsuchiya (IKKT) matrix theory \cite{Sperling:2019xar,Steinacker:2017vqw,Steinacker:2017bhb} (see also e.g. Refs. \cite{Brahma:2021tkh,Hatakeyama:2019jyw,Stern:2014aqa,Chaney:2015ktw,Kim:2011ts,Klinkhamer:2020wct} for related work). These solutions  describe  3+1-dimensional quantum geometries which can be interpreted as 
an effective Friedmann-Lema\^{i}tre-Robertson-Walker (FLRW) cosmology with a BB. In this framework, spacetime along with physical fields  emerge from the basic matrix degrees of freedom and the BB singularity of classical geometry  is completely under control. The study of scalar fields propagating in such a background has been  initiated for the 1+1-dimensional case in Ref. \cite{Karczmarek:2022ejn}  and has been extended to  3+1 dimensions in Ref. \cite{Battista:2022hqn} (the behaviour of fermion fields in a generic curved background provided by the IKKT model has also been investigated, see Ref. \cite{Battista:2022vvl}). 

In the present paper, we provide a concise summary of the main results in Ref. \cite{Battista:2022hqn}
describing the propagation of a  scalar field in an open FLRW bounce-type quantum spacetime in  the framework of the IKKT matrix theory. After reviewing the background in Sec. \ref{Sec:background-geom}, in the first part of the paper we undertake a classical analysis where null and timelike geodesics are studied by exploiting techniques of general relativity (see Sec. \ref{Sec:background-geom}); the quantum aspects are considered in the second part of the paper, where we exploit quantum-field-theory tools to evaluate the scalar field propagator (see Sec. \ref{Sec:Quantum-Analysis}). Finally, we draw our conclusions in Sec. \ref{Sec:Conclusions}.

 \section{The background geometry} \label{Sec:background-geom}

In matrix models, a matrix configuration is a collection of $D$ hermitian matrices $X^a \in {\rm End}(\mathcal{H})$, where $a=1,\dots, D$ and $\mathcal{H}$ is a separable Hilbert space. The matrices $X^a$ can be viewed as  quantized embedding functions
\begin{align} \label{embedding-map-x-X}
X^a \sim x^a : \mathcal{M} \hookrightarrow R^D,    
\end{align}
where $x^a$ are the Cartesian coordinate functions on target space $R^D$ pulled back to $\mathcal{M}$. This means that  the matrices $X^a$ should be viewed as quantizations of the functions $x^a \in \mathcal{C}(\mathcal{M})$. This is indicated in the above equation with the symbol $\sim$,  which means \qm{semi-classical limit}. The matrices $X^a$ generate a noncommutative algebra  which is interpreted as quantized algebra of functions on $\mathcal{M}$. In the semi-classical limit, $\mathcal{M}$ carries a Poisson structure $\{x^a,x^b\} \sim -i [X^a, X^b]$. The embedding map \eqref{embedding-map-x-X} also induces a metric structure on $\mathcal{M}$ via the pull-back of the metric in target space $R^D$. 

In this paper, we consider the spacetime  $\cM^{3,1}$  which can be described in the semi-classical limit as the  projection of fuzzy $H^4_n$   (see Ref. \cite{Sperling:2019xar} for details). Bearing in mind Eq. \eqref{embedding-map-x-X},  this is  obtained from
\begin{align}
   x^a: \quad H^4 \hookrightarrow \R^{4,1}
\end{align}
where $a=0,...,4$. The 4-dimensional hyperboloid can be parametrized as follows
\begin{align}
 \begin{bmatrix}
  x^0 \\ x^1 \\ x^2 \\x^3 \\ x^4
 \end{bmatrix}
= R \begin{bmatrix}
 \cosh(\eta) 
\begin{pmatrix}
\cosh(\chi) \\
\sinh(\chi)\sin(\theta) \cos(\varphi) \\
\sinh(\chi)\sin(\theta) \sin(\varphi) \\
\sinh(\chi)\cos(\theta)
\end{pmatrix} \\
\sinh(\eta) 
\end{bmatrix}, \ 
\label{embedding-4d-hyperboloid}
\end{align}   
where $\eta\in\R$ and  $\chi$ can be restricted to be positive.
Projecting this along the $x^4$ axis leads to a 2-sheeted cover of
the following region
\begin{align}
    x_\mu x^\mu \leq -R^2,
\end{align}
where the upper sheet or \qm{post-BB}  is
covered by $\eta > 0$, while the lower sheet or \qm{pre-BB}  is covered by $\eta < 0$.
The BB separates these sheets, and corresponds to $x_\mu x^\mu = -R^2$.
This leads to the following parametrization of $\cM^{3,1}$
\begin{align}
 \begin{pmatrix}
  x^0 \\ x^1 \\ x^2 \\x^3 
 \end{pmatrix}
= R \cosh(\eta) 
\begin{pmatrix}
\cosh(\chi) \\
\sinh(\chi)\sin(\theta) \cos(\varphi) \\
\sinh(\chi)\sin(\theta) \sin(\varphi) \\
\sinh(\chi)\cos(\theta)
\end{pmatrix} \ .
\label{embedding-3d-hyperboloid}
\end{align}

As shown in Ref. \cite{Sperling:2019xar},  the effective metric on  $\cM^{3,1}$ is the $SO(3,1)$-invariant FLRW metric 
\begin{align}
 d s^2_G = G_{\mu\nu} d x^\mu d x^\nu 
   &= -R^2 \vert \sinh(\eta) \vert^3 d \eta^2 + R^2 \vert \sinh(\eta)\vert \cosh^2(\eta)\, d \Sigma^2 \ \nn\\
   &= -d t^2 + a^2(t)d\Sigma^2 \, ,
   \label{eff-metric-FRW}
\end{align}
where 
\begin{align}
    d\Sigma^2 = d\chi^2 + \sinh^2\chi (d\theta^2 + \sin^2 \theta d\varphi^2),
    \label{dSigma2}
\end{align}
is the invariant length element on the space-like hyperboloids $H^3$. From Eq. \eqref{eff-metric-FRW}, we obtain the form of  the cosmic scale parameter $a(\eta)$ and the relation linking the differentials $dt$ and $d\eta$, i.e.,
\begin{align}
\vert a(\eta) \vert &=  R \cosh(\eta) \vert \sinh(\eta)\vert^{1/2}  ,  
\label{a-eta}
\\
d t &=  R \vert \sinh(\eta)\vert^{3/2} d\eta.  
\label{dt-squared}
\end{align}

\section{Classical analysis: the  behaviour of null and timelike geodesics}\label{Sec:Classical-Analysis}

Before studying the behaviour of null and timelike geodesics, it is worth mentioning that the spacetime geometry \eqref{eff-metric-FRW} possesses a curvature singularity at $\eta=0$, as the analysis of the curvature invariant shows (see Ref. \cite{Battista:2022hqn} for details). For instance, the  Kretschmann scalar reads as 
\begin{equation}\label{kretschmann-invariant}
    R_{\mu \nu \rho \sigma}R^{\mu \nu \rho \sigma}= \dfrac{3}{32R^4 \sinh^{10}\left(\eta\right) } \left[171-60 \cosh\left(2 \eta\right)+25\cosh\left(4 \eta\right)\right],
\end{equation}
and it is seen to blow up at the BB, i.e., at $\eta=0$. Despite that, we will see that null and timelike geodesics are well-defined at the BB, suggesting the presence of a new type of singularity or  what we have dubbed  \qm{mild singularity} in Ref. \cite{Battista:2022hqn}. 

Let us start with the analysis of null geodesics referring to massless particles whose motion starts at some negative value of $\eta$,  reaches the BB at $\eta=0$, and travels away from it for $\eta>0$. Starting from Eq. \eqref{eff-metric-FRW}, this dynamics  is governed by the differential equation 
\begin{equation}
    \dfrac{d \chi}{d \eta} = \vert \tanh \eta \vert \ ,
\end{equation}
which, with the boundary condition $\chi(\eta=0)=0$, leads to 
\begin{equation}
    \chi(\eta) = \left \{ 
\begin{array}{rl}
& \;\;\;   \log \left(\cosh \eta\right),  \quad \, \eta \geq 0,\\
& -\log \left(\cosh \eta\right), \, \quad \eta<0.
\end{array}
\right.
\label{chi-of-eta-solution}
\end{equation}
It is thus clear that null geodesics are continuous at $\eta=0$ and hence light  is able to
travel across the BB (see  Figs. 1 and 2 in Ref. \cite{Battista:2022hqn}). The same conclusions are valid for timelike geodesics, which can be parametrized by
\begin{align}\label{timelike-geod-equation}
    \dfrac{d \chi(\eta)}{d\eta} &=\dfrac{\vert \tanh \eta \vert }{\sqrt{1+a^2(\eta)/\Pi^2}}\ ,
\end{align}
where $\Pi$ is the conserved momentum associated to the  $\chi$-translational Killing vector field underlying the geometry \eqref{eff-metric-FRW}. The numerical analysis of the solution of the above equation reveals that timelike geodesics are well-behaved at the BB (see Fig. 4 in Ref. \cite{Battista:2022hqn}).

\section{Quantum analysis: the scalar field propagator in the IKKT matrix model}\label{Sec:Quantum-Analysis}

In this second part of the paper, we  perform a quantum analysis by  computing the propagator of a scalar field evolving along the FLRW   background \eqref{eff-metric-FRW}. The starting point of our study will be the 2-point function defined by a Gaussian integral in the matrix model, i.e., 
\begin{align}
\langle\phi(x) \phi(y)\rangle 
= \int dk \, \langle \phi_k(x) \phi_k(y)\rangle, 
\end{align}
where
\begin{align}
 \langle \phi_k(x) \phi_k(y)\rangle
= \frac 1Z \int d\phi \,  \phi_k(x) \phi_k(y) e^{i S[\phi_k]},
\end{align}
$Z$ being the generating functional and $S$ the action functional. The necessary details for this computation  will be provided below.
As we will see, some interesting effects due to the presence of the BB at $\eta=0$ will emerge.

\subsection{Eigenfunctions of the  d'Alembertian operator $\Box$}\label{Sec:Eigenfunctions-of-Box}

The \qm{matrix} d'Alembertian governing the propagation of a scalar fields $\phi$ is given by \cite{Battista:2022hqn}
 \begin{align}
    \Box \phi &= \dfrac{1}{R^2}\Biggl[ 3 \tanh (\eta) \partial_\eta  + \partial^2_\eta  -\tanh^2 \eta \left( \dfrac{2}{\tanh \chi}\partial_\chi  +\partial^2_\chi  \right)
    \nonumber \\
   & -\dfrac{\tanh^2 \eta}{\sinh^2 \chi} \left(\dfrac{1}{\tan \theta}\partial_\theta +\partial^2_\theta  +\dfrac{1}{\sin^2 \theta } \partial^2_\varphi \right) \Biggr]\phi,
   \label{Box-expression}
\end{align}
and its eigenfunctions are defined by the equation
\begin{align}
\Box \phi = \lambda \phi.
\label{eigenvalue-problem-BoxG}
\end{align}
If we solve  this equation via the separation \emph{ansatz}
\begin{subequations}
\begin{align}
    \phi(\eta,\chi,\theta,\varphi) &= \tilde{\phi}(\eta,\chi)Y^m_l(\theta,\varphi),
   \\
    \tilde{\phi}(\eta,\chi)&=f(\eta)g(\chi),
\end{align}
\label{separation-ansatz-BoxG}
\end{subequations}
 $Y^m_l(\theta,\varphi)$ being the spherical harmonic functions  of degree $l$ and order $m$ (with $l \geq \vert m \vert$), we end up with the ordinary differential equations
\begin{subequations}
\label{ode-eta&chi-BoxG-operator}
\begin{align}
\left( \partial^2_\eta + 3 \tanh (\eta) \partial_\eta  -\beta  \tanh^2 \eta-\lambda R^2 \right)f(\eta)&=0,
\label{ode-eta-BoxG}
\\
\left(\partial^2_\chi +\dfrac{2}{\tanh \chi} \partial_\chi -\dfrac{l(l+1)}{\sinh^2 \chi}  - \beta\right)g(\chi) &=0,
\label{ode-chi-BoxG}
\end{align}
\end{subequations}
whose solutions, with the appropriate boundary conditions, are
\begin{align}
f(\eta) &= (1-\tanh^2 \eta)^{3/4}\left[ c_1 \mathsf{P}^\mu_\nu(\tanh \eta) \right],
\label{f-eta-solution-BoxG}   
\\
g(\chi)&=\sqrt{\coth^2 \chi -1}  \left[c_2  \mathcal{Q}^{\tilde{\mu}}_l(\coth \chi) \right].  
\label{g-chi-solution-Legendre-BoxG}
\end{align}
In the above equations, $\beta$ is a real-valued constant, $c_1$,  and $c_2$ integration constants,  $\mathsf{P}^{\mu}_\nu(x)$   the associated Legendre function of  the  first kind  with $x$ lying in the interval $(-1,1)$, and $ \mathcal{Q}^{\mu}_\nu(x)$  the associated Legendre function of  the second kind with $x \in (1,+\infty)$; moreover, 
\begin{align}
\nu &= \dfrac{1}{2}\left(2\sqrt{1+\beta}-1\right),
\label{degree-nu-BoxG}
\\
\mu &= \dfrac{1}{2}\sqrt{9+4\beta+4\lambda R^2},
\label{order-mu-BoxG}
\\
\tilde{\mu} &=  \sqrt{1+\beta}.
\label{tilde-mu}
\end{align}

In order to have oscillatory (square-integrable) solutions, we  suppose that both the order \eqref{order-mu-BoxG} of the solution  \eqref{f-eta-solution-BoxG} and the order \eqref{tilde-mu} of the solution \eqref{g-chi-solution-Legendre-BoxG} are  purely imaginary, i.e., 
\begin{align}
\mu &= \pm is,
\label{mu-equals-i-s}
\\
 \tilde{\mu} &=  i q \ ,
\end{align}
where
\begin{align}
s&= \sqrt{-\left(\dfrac{9}{4}+\beta+\lambda R^2\right)} \ >0,
\label{s-equals-vert-mu}    
\\
   q^2 &= -\left(1+\beta\right) >0.
\label{q-equals-vert-tilde-mu}  
\end{align}
The last equation implies that the degree \eqref{degree-nu-BoxG} of the solution \eqref{f-eta-solution-BoxG} is complex and we assume 
\begin{align}
\nu= -\dfrac{1}{2} + i|q|.
\label{degree-nu-with-plus}
\end{align}
Bearing in mind the above equations, the eigenmodes \eqref{separation-ansatz-BoxG} of the d'Alembertian  operator \eqref{Box-expression} having the appropriate boundary conditions are
\begin{align}
\Upsilon^{s_\pm,q}_{l,m}\left(\eta,\chi,\theta,\varphi\right):=\dfrac{1}{\sqrt{\cosh^3 \eta} \sinh \chi }\mathsf{P}^{\pm is}_\nu \left(\tanh \eta \right) \mathcal{Q}^{iq}_l \left(\coth \chi\right) Y^m_l(\theta,\varphi), \qquad q \in\R,  s>0,
\label{Upsilon}    
\end{align}
where we have assumed $\chi>0$. 

As shown in details in Ref. \cite{Battista:2022hqn}, the above eigenfunctions satisfy the following orthogonality relations:
\begin{align}
\langle \Upsilon^{s^\prime_{+},q^\prime}_{l^\prime,m^\prime}, \Upsilon^{s_{+},q}_{l,m}\rangle 
&=\dfrac{e^{-2 \pi q}\left(\pi/2\right)^2}{q \sinh \left(\pi q\right)}  \delta_{l l^\prime} \delta_{m m^\prime} \delta(q-q^\prime) \Biggl[a(q,s)\delta(s+s^\prime) + b(q,s)\delta(s-s^\prime)\Biggr],
\label{orthogonality-relations-final}
\end{align}
where 
\begin{subequations}
\label{a-q-s-and-b-q-s}
\begin{align}
a(q,s) &= \dfrac{2 \pi \cosh(\pi q)}{s \sinh(\pi s)}\dfrac{1}{\Gamma\left(iq-is+1/2\right)\Gamma(-iq-is+1/2)}=a(q,-s)^*,
\\
b(q,s) &= \dfrac{2 \sinh(\pi s)}{s}\left[1+\dfrac{\cosh^2(\pi q)}{\sinh^2(\pi s)}\right]=b(q,-s),
\end{align}
\end{subequations}
$\Gamma(x)$ being the  gamma function and $\delta(x)$ the Dirac-delta function.

For future purposes, it will be important to consider the following 
\qm{flat} regime, denoted with \qm{FR}: 
\begin{align}
\mbox{FR}:\quad  \chi  < 1, \qquad q \gg l,
\label{regime-A}
\end{align}
where  $q$ will be a typical momentum. In this regime and  for large times (i.e., $\eta \to + \infty$) the eigenmodes become
\begin{align}
\Upsilon^{s_{\pm},q}_{l,m}(\eta,\chi,\theta,\varphi)&\overunderset{\eta \to + \infty}{{\rm FR}}{\sim}\, \dfrac{1}{\sqrt{\cosh^3 \eta}}\dfrac{e^{-\pi q} j_l\left(q \chi\right) \Gamma(iq+l+1)}{ q^l} \dfrac{e^{\pm i \eta s}}{\Gamma(1\mp is)}Y^{m}_l(\theta,\varphi),
\label{flate-regime-Upsilon}
\end{align}
where  $ j_l(x)$ are the spherical Bessel functions.

\subsection{The propagator}

In order to calculate the propagator of a scalar field $\phi$ having mass $m$, we recall that its action reads as, in the semi-classical limit,
\begin{align}
S_{\varepsilon}\left[\phi\right]= \int \Omega   \phi^*(x)\left(-\Box -m^2 + i \varepsilon \right)\phi(x), \label{action-1}
\end{align}
where
\begin{align}
  \Omega &= \cosh^3(\eta) d\eta\sinh^2(\chi) d\chi \sin(\theta)d\theta d\varphi
  \label{symplectic-volume-form}
\end{align}
is the $SO(4,1)$-invariant  volume form on $H^4$. If we decompose $\phi(x)$ in the basis of the eigenmodes \eqref{Upsilon} as follows 
\begin{align}
\phi(x)=  \sum_{l,m}\int ds dq \Bigl[\phi^+ \Upsilon^{s_{+},q}_{l,m}(x)+\phi^- \Upsilon^{s_{-},q}_{l,m}(x)  \Bigr],
\label{phi-decomposition}
\end{align}
then  we can write
\begin{align}
     S_{\varepsilon}\left[\phi\right] &=   \sum_{l,m}\sum_{l^\prime,m^\prime}\int   ds dq ds^\prime dq^\prime  \left[-\dfrac{1}{R^2}\left(q^2-s^2-\dfrac{5}{4}\right)-m^2 + i \varepsilon \right] \dfrac{e^{-2\pi q}(\pi/2)^2 }{q \sinh(\pi q)} 
\nonumber \\
&\times \delta_{l l^\prime} \delta_{m m^\prime} \delta(q-q^\prime) \delta(s-s^\prime)  \begin{bmatrix}
 \left(\phi^{\prime +}\right)^* & \left(\phi^{\prime -}\right)^*
\end{bmatrix}
\mathscr{B}(q,s)  \begin{bmatrix}
 \phi^+ \\ 
 \phi^-
\end{bmatrix},
\label{action-final-expression}
\end{align}
where we have exploited Eq. \eqref{orthogonality-relations-final} and  
\begin{align}
\mathscr{B}(q,s)=
\begin{bmatrix}
b(q,s) & a(q,-s)\\
a(q,s) & b(q,-s)
\end{bmatrix}.
\end{align}
Therefore, the propagator in momentum space reads as
\begin{align}
\left \langle \left(\Phi^{\pm}\right)   \left( \Phi^{\prime \pm}\right)^{\dagger} \right \rangle &= \delta_{l l^\prime} \delta_{m m^\prime} \delta\left(q-q^\prime\right)
\delta\left(s-s^\prime\right) \dfrac{1}{\dfrac{1}{R^2}\left(s^2 - q^2 +\dfrac{5}{4}\right)-m^2 + i \varepsilon} 
\nonumber \\
& \times \dfrac{4q \sinh (\pi q)}{e^{-2 \pi q}\pi^2} \left[\mathscr{B}(q,s)\right]^{-1},
\label{propagator-momentum-space}
\end{align}
while in position space 
\begin{align}
  \langle \phi(x) \phi^*(x^\prime)\rangle &= \sum_{l,m}\sum_{l^\prime,m^\prime} \int ds dq ds^\prime dq^\prime 
\begin{bmatrix}
 \Upsilon^{s_+,q}_{l,m} (x) & \Upsilon^{s_-,q}_{l,m}(x)
\end{bmatrix}  
  \left \langle \left(\Phi^{\pm}\right)   \left( \Phi^{\prime \pm}\right)^{\dagger} \right \rangle
  \begin{bmatrix}
  \left( \Upsilon^{s^\prime_{+},q^\prime}_{l^\prime,m^\prime} \left(x^\prime\right)\right)^* \\
   \left( \Upsilon^{s^\prime_{-},q^\prime}_{l^\prime,m^\prime} \left(x^\prime\right)\right)^* 
  \end{bmatrix},
\label{propagator-position-space}  
\end{align}
where we have adopted the compact notation 
\begin{align}
\Phi^{\pm} &\equiv \begin{bmatrix}
 \phi^+ \\ 
 \phi^-
\end{bmatrix}.
\end{align}

\subsubsection{The propagator in the flat regime and with $\eta \to + \infty$}

In the flat regime   \eq{regime-A} and when $\eta$ goes to infinity,  the eigenmodes \eqref{Upsilon}  reduce to \eqref{flate-regime-Upsilon}. Therefore, starting from Eq.  \eqref{propagator-position-space}   the  late-time  local propagator can be written as the sum of a leading piece (denoted by \qm{L}) and a subleading part (denoted by \qm{SL}), i.e., 
\begin{align}
\langle \phi(x) \phi^*(x^\prime)\rangle \overunderset{\eta \to + \infty}{{\rm FR}}{\sim} \,  \langle \phi(x) \phi^*(x^\prime)\rangle^{\eta \to + \infty, {\rm FR}}_{\rm L} +\langle \phi(x) \phi^*(x^\prime)\rangle^{\eta \to + \infty, {\rm FR}}_{\rm SL}.
\label{flat-regime-propagator-leading&subleading}     
\end{align}
The leading contribution is \cite{Battista:2022hqn}
\begin{align}
\langle \phi(x) \phi^*(x^\prime)\rangle^{\eta \to + \infty,{\rm FR}}_{\rm L} &=\dfrac{4R^2}{\pi^2}  \sum_{l,m} \dfrac{ Y^m_l(\theta,\varphi)\left[Y^m_l(\theta^\prime,\varphi^\prime)\right]^*}{\sqrt{\left(\cosh^3 \eta\right)\left(\cosh^3 \eta^\prime\right)}} \int^{+ \infty}_{-\infty} ds\, e^{i s\left(\eta - \eta^\prime\right)}
\nonumber \\
& \times \int^{+ \infty}_{0} dq
\dfrac{q^2 j_l\left(q \chi\right)j_l\left(q \chi^\prime\right)}{\left(s^2 - q^2 + \dfrac{5}{4}-m^2 R^2 + i \varepsilon \right)}, 
\label{propagator-flat-regime-leading-2}
\end{align}
whereas the subleading term is 
\begin{align}
\langle \phi(x) \phi^*(x^\prime)\rangle^{\eta \to + \infty,{\rm FR}}_{\rm SL} &=\dfrac{4R^2}{\pi^4}  \sum_{l,m} \dfrac{ Y^m_l(\theta,\varphi)\left[Y^m_l(\theta^\prime,\varphi^\prime)\right]^*}{\sqrt{\left(\cosh^3 \eta\right)\left(\cosh^3 \eta^\prime\right)}} \int^{+ \infty}_{-\infty} ds \, e^{is\left(\eta+\eta^\prime\right)} 
\nn \\
& \times \int^{+ \infty}_{0}dq
\dfrac{j_l\left(q \chi\right)j_l\left(q \chi^\prime\right)q^2 s  \cosh(\pi q)   \sinh(\pi s)}{\left(s^2 - q^2 + \dfrac{5}{4} -m^2 R^2 + i \varepsilon \right)} 
\nonumber \\
& \times  \Gamma\left(\dfrac{1}{2}-iq-is\right)\Gamma\left(\dfrac{1}{2}+iq-is\right)\Gamma^2\left(is\right).
\label{propagator-flat-regime-subleading}
\end{align}

The most important result of this section is that Eq. \eqref{propagator-flat-regime-leading-2} resembles, up to an  $\eta$-dependent normalization factor,  the usual local Feynman propagator on a flat four-dimensional spacetime.  

\subsubsection{The propagator in the flat regime and with $\eta \to 0$} \label{Sec:prop-early-time}

The behaviour of the scalar field near the BB can be understood by considering the features of the propagator for small times, i.e., when $\eta, \eta' \to 0$.  For early times and in the flat regime \eqref{regime-A}, the eigenfunctions \eqref{Upsilon}  become
\begin{align}
\Upsilon^{s_{\pm},q}_{l,m}(\eta,\chi,\theta,\varphi)&\overunderset{\eta \to 0}{{\rm FR}}{\sim}\,\dfrac{\sqrt{\pi}\,2^{\pm is}\, e^{\pm i \eta s} }{\Gamma \left(\dfrac{3}{4}-\dfrac{iq}{2} \mp \dfrac{is}{2}\right)\Gamma \left(\dfrac{3}{4}+\dfrac{iq}{2} \mp \dfrac{is}{2}\right)}\dfrac{e^{-\pi q} j_l\left(q \chi\right) \Gamma(iq+l+1)}{ q^l} Y^m_l \left(\theta,\varphi\right),
\label{upsilon-flat-regime-small-time}
\end{align}
and hence the early-time propagator assumes the form \cite{Battista:2022hqn}
\begin{align}
\langle \phi(x) \phi^*(x^\prime)\rangle\overunderset{\eta,\eta^\prime \to 0}{{\rm FR}}{\sim}& \, 4R^2\sum_{l,m} Y^m_l(\theta,\varphi)\left[Y^m_l(\theta^\prime,\varphi^\prime)\right]^* \int\limits^{+ \infty}_{- \infty} ds     \int\limits^{+ \infty}_{0} dq\,    \dfrac{ q^2 j_l \left( q \chi\right)j_l \left( q \chi^\prime \right) e^{is \left(\eta - \eta^\prime \right)}  }{ \left(s^2 - q^2 + \dfrac{5}{4} -m^2 R^2 + i \varepsilon \right)}
\nn \\
& \times
\dfrac{- is}{\left[\cosh( \pi q)-i\sinh( \pi s)\right]} \dfrac{1}{\left\vert \Gamma\left(\dfrac{3}{4}+\dfrac{iq}{2}+\dfrac{is}{2}\right)\Gamma\left(\dfrac{3}{4}+\dfrac{iq}{2}-\dfrac{is}{2}\right) \right\vert^2 }.
\label{early-time-propagator-final-1}
\end{align}

This propagator leads  to a well-defined  propagation between two points located on opposite sheets of the spacetime $\cM^{3,1}$  near the BB, while for  for larger $|\eta|$ and $|\eta^\prime|$ this is suppressed compared to the case of two points on the same sheet. This means that  the quantum analysis performed in this section agrees,  at least qualitatively, with the  classical investigation of Sec. \ref{Sec:Classical-Analysis}, as we have found again that a scalar field can travel through the BB without hindrance whatsoever.

\section{Conclusions}\label{Sec:Conclusions}

In this paper, we have studied the propagation of a scalar field on a quantum version of a 3+1-dimensional bouncing FLRW spacetime provided by the framework of the IKKT matrix models. The paper has been morally divided into two parts, a classical and a quantum one, which lead to the same conclusion: a scalar field $\phi$ is able to travel across the BB located at $\eta=0$. It should be taken into account that our analysis is restricted to non-interacting test particles on the background geometry. This is of course not entirely satisfactory, due to the singular behaviour of the density of matter near the BB, which would lead to  modifications of the background. The inclusion of these effects along with the induced Einstein-Hilbert action deserves further consideration in a separate paper.

\section*{Acknowledgement}

This work  was supported by the Austrian Science Fund (FWF) grant P32086.

\bibliographystyle{JHEP}
\bibliography{references}

\end{document}